\documentclass[twocolumn,showpacs,preprintnumbers,amsmath,amssymb,showkeys]{revtex4}
\usepackage{graphicx}
\usepackage{dcolumn}
\usepackage{bm}

\begin{document}




\title{Transport, magnetic and superconducting properties of RuSr$_2$$R$Cu$_2$O$_8$ 
($R=$ Eu, Gd) doped with Sn}

\author{M.~Po\v{z}ek}
\author{I.~Kup\v{c}i\'{c}}
\author{A.~Dul\v{c}i\'{c}}
\author{A.~Hamzi\'{c}}
\author{D.~Paar}
\author{M.~Basleti\'{c}}
\author{E.~Tafra} 
\affiliation{Department of Physics, Faculty of Science, University of Zagreb, 
P. O. Box 331, HR-10002 Zagreb, Croatia}
\author{G.~V.~M.~Williams}
\affiliation{Industrial Research Limited, P.O. Box 31310, 
Lower Hutt, New Zealand}

\begin{abstract}
Ru$_{1-x}$Sn$_x$Sr$_2$EuCu$_2$O$_8$ and Ru$_{1-x}$Sn$_x$Sr$_2$GdCu$_2$O$_8$ 
have been comprehensively studied by microwave and dc resistivity and 
magnetoresistivity and by the dc Hall measurements. The magnetic ordering 
temperature $T_m$ is considerably reduced with increasing Sn content. 
However, doping with Sn leads to only slight reduction of the superconducting 
critical temperature $T_c$ accompanied with the increase of the upper 
critical field $B_{c2}$, indicating an increased disorder in the system 
and a reduced scattering length of the conducting holes in CuO$_2$ layers. 
In spite of the increased scattering rate, the normal state resistivity 
and the Hall resistivity are reduced with respect to the pure compound, 
due to the increased number of itinerant holes in CuO$_2$ layers,  which 
represent the main conductivity channel. Most of the electrons in RuO$_2$ 
layers are presumably localized, but the observed negative magnetoresistance and the 
extraordinary Hall effect lead to the conclusion that there exists a 
small number of itinerant electrons in RuO$_2$ layers that exhibit 
colossal magnetoresistance.
\end{abstract}


\maketitle

\section{Introduction}
%
\label{sec:level1}

Despite the intensive investigation of ruthenate-cuprates
in the last ten years,  a lot of questions about the nature 
of magnetic order and superconductivity, in particular their coexistence, 
are still unanswered \cite{Nachtrab:06}.
It is well established that in RuSr$_2$$R$Cu$_2$O$_8$ 
(Ru1212$R$, $R=$ Eu, Gd, Y)  compounds the magnetic ordering of 
ruthenium sublattice occurs at about 130 K. 
The ordering is predominantly antiferromagnetic (AFM) with an easy 
axis perpendicular to the layers and  with a weak ferromagnetic (FM) 
component parallel to the layers. 

While both, RuO$_2$ and CuO$_2$ layers may
participate in the normal-state conductivity, 
the superconducting (SC) properties of Ru1212$R$ compounds 
are associated only with the charge carriers in  the CuO$_2$ planes, 
and are, thus,  strongly dependent on the concentration of these carriers. 
The related superconducting critical temperature $T_c$  ranges between 15 and 50 K, 
depending on the sample composition and/or sample preparation conditions.
According to  the thermopower and Hall coefficient measurements, 
pure Ru1212$R$ samples are expected to be intrinsically 
underdoped, with the  effective hole concentration in the CuO$_2$ 
planes nearly equal to $p\approx 0.07$ \cite{McCrone:03}. 

In  high-$T_c$  superconductors, SC properties 
are strongly dependent on the effective hole concentration $p$
in the CuO$_2$ planes. 
In the optimally doped systems, the effective hole concentration 
is estimated to be $p\approx 0.16$ holes/Cu.
In this respect, in recent years there has been a lot of experimental  
effort to adjust the number of charge carriers in the CuO$_2$ planes  
in Ru1212$R$ samples using  different substitutional impurities.
The largest changes in $T_c$ are found when Ru is partially replaced by Cu
($T_c \approx 75$ K for  40\% of Ru replaced by Cu \cite{Klamut:01}).
In several studies ruthenium was replaced by Nb$^{5+}$,V$^{4+,5+}$ Sn$^{4+}$, Ti$^{4+}$ and Rh$^{3+}$ 
\cite{McCrone:03,McLaughlin:01,Williams:03,Hassen:06,McLaughlin:99,Malo:00,Hassen:03,Yamada:04,Steiger:07}. 
Substitutions have also been made for other ions.
For example, trivalent gadolinium was replaced by Ca$^{2+}$ and Ce$^{4+}$ \cite{Klamut:01a} 
or by isovalent Y and Dy \cite{McCrone:03}. 
Finally, divalent strontium was replaced by La$^{3+}$ and Na$^{+}$ 
\cite{Williams:03,Hassen:06}. 
If the substitutional impurity lowers the total number of holes
(La$^{3+}$ for Sr$^{2+}$, Ce$^{4+}$ for Gd$^{3+}$, Nb$^{5+}$ for Ru$^{4+}$), $T_c$ is strongly depressed,
indicating that $p$ is reduced well below $0.07$. 
However, in the cases when valence counting of impurities
would suggest a strong increase in $p$ 
(Na$^{+}$ for Sr$^{2+}$, Sn$^{4+}$ for Ru$^{5+}$), 
the increase in $T_c$ was not experimentally observed, contrary to the observation in 
common high-$T_c$ systems with substitutional impurities.
This is rather similar to the observation in the underdoped YBa$_2$Cu$_3$O$_{7-x}$
compounds where $T_c \approx 60\rm \, K$ is nearly constant in a wide range of $x$. 
The increase of $T_c$ was only observed in Ca$^{2+}$ substitution for Gd$^{3+}$ \cite{Klamut:01a},
but it can be attributed to the formation of Ru$_{1-x}$Sr$_2$GdCu$_{2+x}$O$_8$ phase.
The magnetic ordering is, however, strongly influenced by the substitutions in all 
of the mentioned doping studies.

There is still an open debate whether RuO$_2$ layers are conducting or not. 
The magnetization \cite{Papageorgiou:07} and NQR \cite{Tokunaga:01,Kumagai:01} results
suggest that most of the electrons are localized resulting in Ru$^{4+}$ ions 
while magnetoresistivity and Hall 
measurements \cite{McCrone:03,Pozek:02} suggest the existence of conductivity 
in magnetically ordered RuO$_2$ layers.
We shall deal with this question in our doping study, where the number of charge carriers
is modified by the replacement of Ru ions with Sn.
In this paper a comprehensive microwave study on several Ru1212Eu 
and Ru1212Gd samples doped with Sn is reported. 
One of the samples (Ru$_{0.8}$Sn$_{0.2}$Sr$_2$GdCu$_2$O$_8$) 
is also characterized by transport, magnetoresistance and Hall measurements, and the results 
are compared with our previous measurements \cite{Pozek:02,Pozek:07}. 

In this context, it is important to point out the controversy related to the role 
of the Sn substitution on  Ru sites in Ru1212Gd samples.
McLaughlin et al. \cite{McLaughlin:01,McLaughlin:99} 
and Hassen and Mandal \cite{Hassen:06} reported, respectively, 
the increase and decrease of $T_c$ with doping. 
It is possible that this controversy reflects different ways 
to experimentally determine $T_c$, or might be related to different 
methods of sample preparation.
Moreover, we shall revise the observation in the Hall study (Fig. 2 in Ref. \cite{McCrone:03}) 
in which the increase of the Hall coefficient with increased Sn content was reported.

\section{Experimental details}
%

Ru$_{1-x}$Sn$_x$Sr$_2$$R$Cu$_2$O$_8$ 
polycrystalline samples, where $R=$Eu or Gd, were prepared by solid state synthesis, 
as described elsewhere \cite{Williams:03}.
 
An elliptical $_e$TE$_{111}$ copper cavity operating at 9.3~GHz was used for the microwave 
measurements. The sample was placed in the center of the cavity on a sapphire holder. 
At this position microwave electric field has its maximum. 
External dc magnetic field perpendicular to the microwave electric field was varied 
from zero up to 8 T. 
The temperature of the sample could be varied from liquid helium to room temperature.  
The measured quantity was  $1/2Q$, the total losses of the cavity loaded by the sample. 
It is simply related to the surface resistance of the material $R_s$ which comprises both, 
nonresonant resistance and resonant spin contributions. 
The details of the detection scheme are given elsewhere \cite{Nebendahl:01}.  

Resistivity, magnetoresistance and Hall effect measurements were carried out in the standard 
six-contact configuration using the rotational sample holder and the conventional ac 
technique (22~Hz, 1~mA), in magnetic fields up to 8 T. 
Temperature sweeps for the resistivity measurements were performed with carbon-glass 
and platinum thermometers, while magnetic field dependent sweeps were done at constant 
temperatures which was controlled with a capacitance thermometer.

\section{Results and analyses}
%

\begin{figure}[t]
\includegraphics[width=16pc]{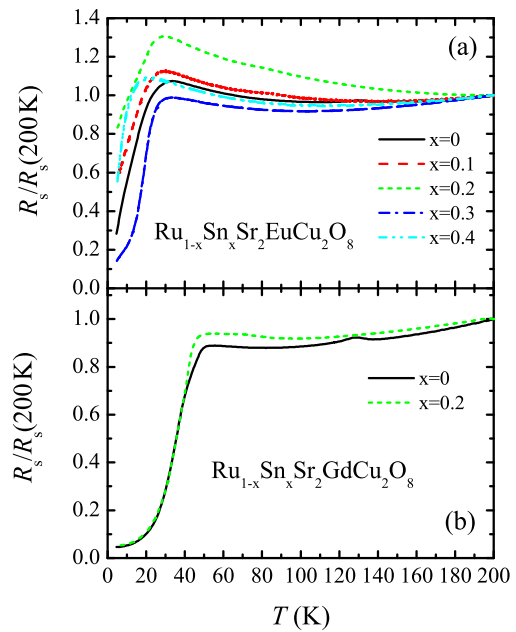}
\caption{(color online) Temperature dependence of the normalized surface resistance of 
(a) Ru$_{1-x}$Sn$_x$Sr$_2$EuCu$_2$O$_8$ and 
(b) Ru$_{1-x}$Sn$_x$Sr$_2$GdCu$_2$O$_8$ for various Sn concentrations $x$.} 
\label{Rs_normalized}
\end{figure}

\begin{figure}[t]
\includegraphics[width=16pc]{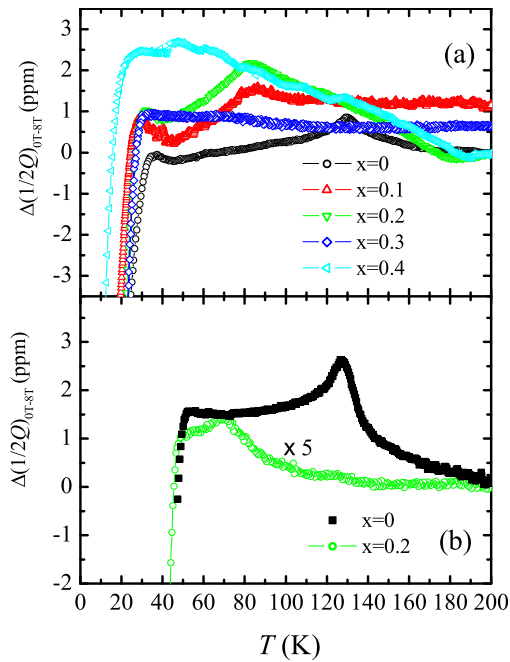}
\caption{(color online) Differences in the microwave resistance in zero field and $B=8$ T of 
(a) Ru$_{1-x}$Sn$_x$Sr$_2$EuCu$_2$O$_8$
and (b) Ru$_{1-x}$Sn$_x$Sr$_2$GdCu$_2$O$_8$ for various $x$.} 
\label{Rs_differences}
\end{figure}

The temperature dependence of the microwave surface resistance 
of various Ru1212Eu (Sn doping $x = 0$, 0.1, 0.2, 0.3 and 0.4) 
and Ru1212Gd ($x = 0$, 0.2) compounds is shown
in Figs.~\ref{Rs_normalized}(a) and  \ref{Rs_normalized}(b), respectively.
The data are normalized at $T=200$ K, for comparison. 
We found  that, for the samples of similar geometry,
the absolute level of the normal-state absorption in the Eu-based samples 
was systematically higher than the absorption in the Gd-based samples.
Moreover, the surface resistance of the Eu-based samples shows more pronounced
rise as the temperature decreases towards $T_c^{\rm onset}$
($T_c$, hereafter) than it is the case in Gd-based samples.
The related crossover to the SC state occurs at lower temperatures.
The SC crossover temperature region is rather broad in these data, 
resulting in a rather uncertain estimation of $T_c$. 
This problem can be easily solved by taking into account that the 
superconductivity in the vicinity of $T_c$ is strongly dependent 
on magnetic fields.
For this purpose,  the difference between the microwave absorption 
in zero field and in $B=8$ T is plotted in Fig.~\ref{Rs_differences}. 
[A considerably smaller signal of Sn-doped sample in Figs.~\ref{Rs_differences}(b) and 
\ref{Rs_magnetic}(d), with respect to the pure Gd-based sample, is only due to the 
smaller geometry of the sample.]
Both the magnetic and SC transitions can be detected  here
with a much better resolution than in Fig.~1.


\begin{figure}[t]
\includegraphics[width=20pc]{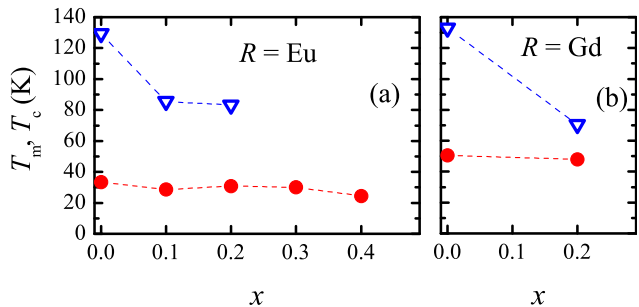}
\caption{(color online) Dependence of  $T_m$ (open triangles) and  $T_c$ (full circles) 
on $x$ in Ru$_{1-x}$Sn$_{x}$Sr$_2 M$Cu$_2$O$_8$, for 
(a) $R=$ Eu  and (b) $R=$ Gd.} 
\label{phase}
\end{figure}

\begin{figure}[t]
\includegraphics[width=18pc]{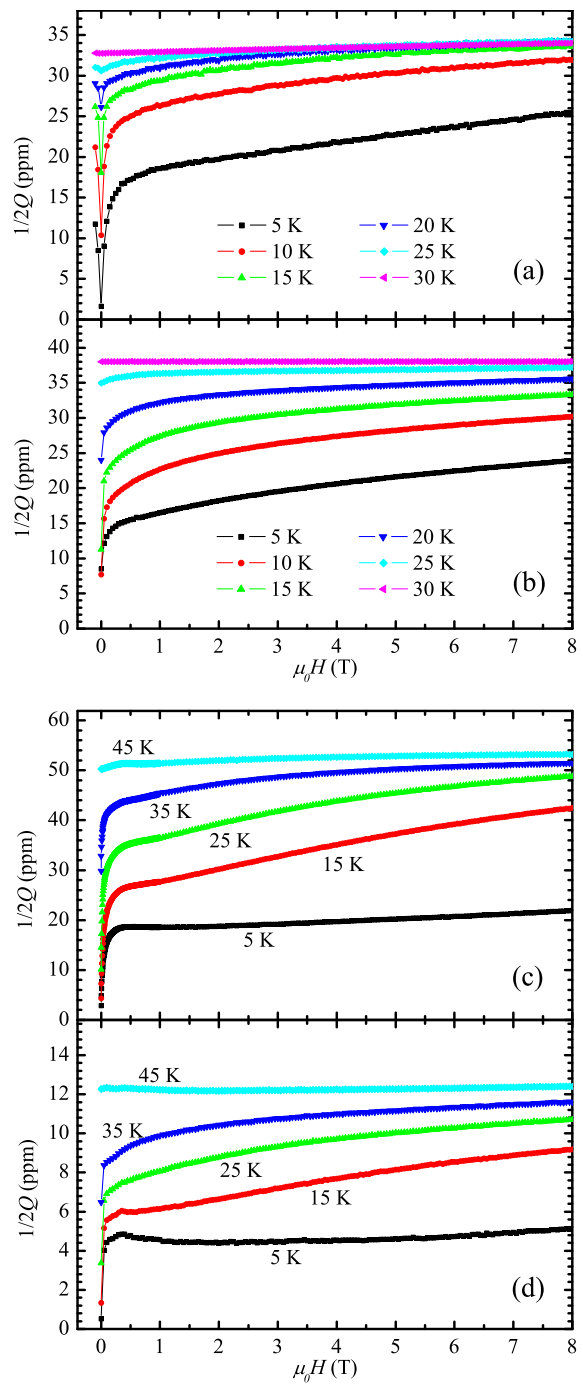}
\caption{(color online) Magnetic field dependence of the microwave absorption 
for (a) RuSr$_2$EuCu$_2$O$_8$, 
(b) Ru$_{0.7}$Sn$_{0.3}$Sr$_2$EuCu$_2$O$_8$, (c) RuSr$_2$GdCu$_2$O$_8$ and
(d) Ru$_{0.8}$Sn$_{0.2}$Sr$_2$GdCu$_2$O$_8$ measured at various temperatures.} 
\label{Rs_magnetic}
\end{figure}

The magnetic ordering temperature $T_m$
of the Ru lattice corresponds with small peaks 
clearly seen in Fig.~\ref{Rs_differences}.
The dependence of  $T_m$ on the Sn content $x$ is shown in Figs.~\ref{phase}(a)
and \ref{phase}(b). 
$T_m$ is strongly suppressed with increasing $x$ in both the Gd- 
and Eu-based samples 
(the magnetic critical
temperature $T_m$ is reduced nearly by a factor of 2 for $\Delta x \approx 0.2$).
This suppression seems to be a consequence of the reduced content of 
the Ru magnetic ions (dilution of the magnetic lattice)
and the increased disorder in the Ru lattice.
Similar decrease in $T_m$ is found in the samples in which Ru is 
replaced by nonmagnetic Nb$^{5+}$ ions \cite{McCrone:03}
($\Delta T_m \approx 30$ K for $\Delta x \approx 0.2$ \cite{McLaughlin:01}).

The SC ordering temperature $T_c$
is clearly seen in Fig.~\ref{Rs_differences} for all the samples,
and its dependence on $x$ is further shown in Fig.~\ref{phase}. 
The sudden drop in the difference of the microwave absorption 
in zero field and in $B=8$ T indicates the onset of the SC 
state in the grains.
This effect can be associated with the intrinsic value of $T_c$.
On the other hand, the whole sample is expected to become 
superconducting at much lower temperatures where the 3D network 
of intergranular Josephson junctions is established.
It is clear that the influence of Sn doping 
on $T_c$ is not nearly as  dramatic as the influence on $T_m$. 
In the Gd-based samples, for example, the replacement of 20\% of Ru 
by Sn does not appreciably change $T_c$. 
In the Eu-based samples, on the other hand, a slight reduction of $T_c$ 
with increasing $x$ is observed, but it is negligible in comparison 
with the related change in $T_m$.

The SC state is further analyzed 
by measuring the magnetic field dependence of the microwave absorption. 
The results taken at different temperatures are shown in 
Fig.~\ref{Rs_magnetic} for two undoped samples and two Sn-doped samples. 
At lower fields (below 0.5 T), the microwave absorption 
is strongly field-dependent due to the intergranular superconductivity
dominated by the 3D Josephson network. 
At higher fields, all the Josephson junctions are driven to the normal state, 
and superconductivity is localized only in the grains. 
The microwave absorption grows slowly in this field range as the number of vortices increases, 
and it is expected to reach the normal-state absorption at the upper critical 
field $B_{c2}$. 
\begin{figure*}[tb]
\includegraphics[width=36pc]{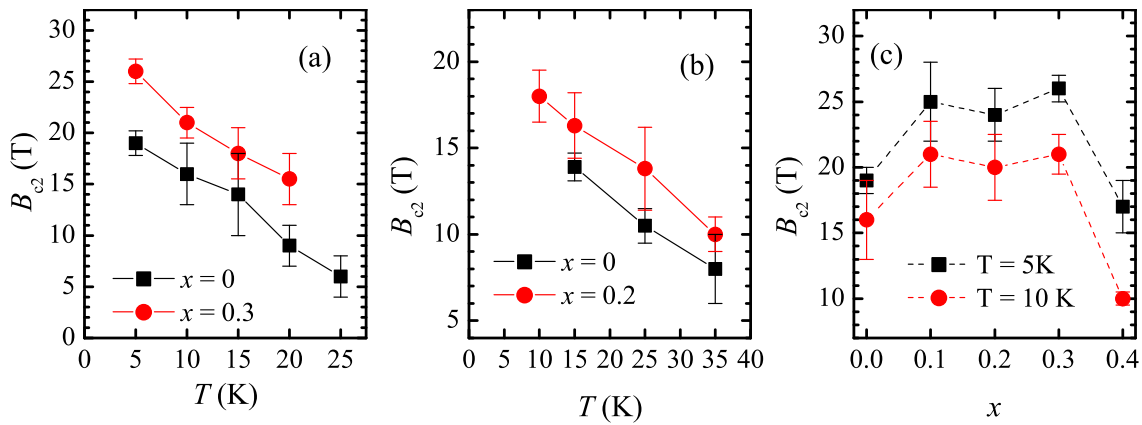}
\caption{(color online) (a) Dependence of the upper critical field $B_{c2}$ on temperature 
in Ru$_{1-x}$Sn$_{x}$Sr$_2$EuCu$_2$O$_8$, for $x=0$ (squares) 
and $x=0.3$ (circles). 
(b) Dependence of $B_{c2}$ on temperature in 
Ru$_{1-x}$Sn$_{x}$Sr$_2$GdCu$_2$O$_8$, for $x=0$ (squares) 
and $x=0.2$ (circles). 
(c) Dependence of $B_{c2}$ on $x$ in Ru$_{1-x}$Sn$_{x}$Sr$_2$EuCu$_2$O$_8$ 
at $T=5$ K (squares) and $T=10$ K (circles).} 
\label{Bc2}
\end{figure*}
The microwave magnetoresistance is therefore a useful tool 
for the estimation of upper critical fields. 
If one assumes that the depinning frequency is lower than 
the driving microwave frequency, it suffices to extrapolate the microwave 
absorption curves to the normal-state values and determine  $B_{c2}$. 
However, this simple determination of $B_{c2}$ 
is not possible in  cases where the depinning frequency  is comparable 
to the driving frequency. 
Then, one has to take into account also the frequency 
shift.
The details of the complete  procedure are explaned in Ref.~\cite{Janjusevic:06}.
The shape of the low-temperature microwave magnetoresistance 
curves of the Gd-based samples [$T=5$ K curves in Figs.~\ref{Rs_magnetic}(c) 
and \ref{Rs_magnetic}(d)] points at the regime where  the two frequencies 
are comparable to each other. 
At higher temperatures, on the other hand, the driving microwave frequency
seems to be larger than the depinning frequency allowing 
the direct estimation of the upper critical fields. 
In the Eu-based samples, the depinning frequency seems to be 
low enough even at low temperatures, so that the upper critical fields 
can be simply determined in the complete temperature region of  
Figs.~\ref{Rs_magnetic}(a)  and \ref{Rs_magnetic}(b).

The upper critical fields $B_{c2}$ estimated from the curves in Fig.~\ref{Rs_magnetic} 
(and similar curves for other samples not shown here) 
are plotted in Fig.~\ref{Bc2}. 
The comparison of $B_{c2}$ estimated in two Eu-based samples, 
$x=0$ and 0.3, is shown  in Fig.~\ref{Bc2}(a). 
The analogous comparison of $B_{c2}$ in the Gd-based samples with 
$x=0$ and 0.2 is given in Fig.~\ref{Bc2}(b). 
From these figures one can see 
that the upper critical fields  in the Gd-based 
samples are lower than in the Eu-based samples, and, in both cases, 
the upper critical fields are higher in the doped samples than in the undoped
samples. 
Having in mind that higher $B_{c2}$ means lower Ginzburg-Landau coherence length, 
which, in turn, depends on the electron mean free path, 
one concludes that the disorder in the undoped Eu-based samples is higher 
than in the undoped Gd-based samples, and that the Sn doping  introduces 
an additional disorder in both systems. 
The low-temperature ($T=5$ and 10 K) dependence 
of the upper critical field on the Sn concentration 
$x$ in the Eu-based samples is also shown [Fig.~\ref{Bc2}(c)]. 
From Fig.~\ref{Bc2}(c), we can see that
the upper critical field increases with $x$ for low to moderate 
concentrations, while for the highest concentration ($x=0.4$) 
$B_{c2}$ is suppressed. 
The drop of $B_{c2}$ in the $x=0.4$ sample  
is related to the drop of $T_c$ in Fig.~\ref{phase}(a).

In order to clarify the role of the disorder introduced by Sn doping,
we extend the experimental study and analysis to transport coefficients.
We have chosen the doped sample 
Ru$_{0.8}$Sn$_{0.2}$Sr$_2$GdCu$_2$O$_8$ which exhibits a sharp SC onset in 
Fig.~\ref{Rs_differences}(b), and  measured 
its resistivity, magnetoresistivity  and Hall resistivity. 

\begin{figure}[tb]
\includegraphics[width=18pc]{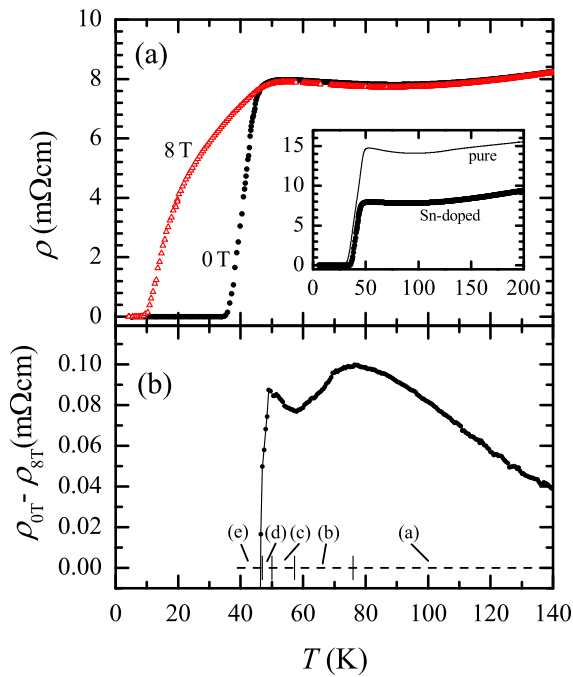}
\caption{(color online) (a) Resistivity of Ru$_{0.8}$Sn$_{0.2}$Sr$_2$GdCu$_2$O$_8$ 
measured in zero field (black line) and in $B=8$ T (red line). 
Inset shows the comparison of resistivity with the resistivity 
of the undoped sample (thin line) \cite{Pozek:02}.
(b) Difference between the  zero-field and  8 T resistivities. 
Five temperature intervals labeled by (a), (b), (c), (d), and (e)
correspond to subsets of MR curves in Fig.~\ref{MRall}.
} 
\label{rho}
\end{figure}

Figure \ref{rho}(a) shows the resistivity measured in zero magnetic field 
and in $\mu_0 H=8\rm \, T$. 
In both cases, the resistivity at 200 K is 9.5 m$\Omega$cm, which is 
much lower than 15.5 m$\Omega$cm of the undoped sample \cite{Pozek:02}. 
In zero field, one observes a relatively broad SC transition 
characterized by the change in the magnitude from 90\% to 10\% 
within the temperature interval $\Delta T = 8$ K. 
Such temperature dependence of the resisitivity in the vicinity of $T_c$ 
is usually attributed to the spontaneous vortex phase. 
The temperature interval characterizing 
the SC transition  in the 8 T resistivity  data is much broader than that 
of the zero-field data, and 
 the zero resistance occurs only at 10 K, 
indicating that the pinning above 10 K is not strong 
enough to suppress the vortex motion, even for dc driving currents. 
The absence of strong pinning forces above 10 K justifies the  estimation 
of $B_{c2}$ from the magnetic field dependence 
of the  microwave absorption given above [Fig.~\ref{Rs_magnetic}(d)].

The onset of superconductivity in the grains could be seen better  
if the  resistivity in $8 \rm \, T$ is subtracted from the zero-field resistivity,
as shown in Fig.~\ref{rho}(b).
The subtraction also reveals the existence of 
a broad maximum at the magnetic ordering temperature 
$T \approx T_{m} \approx 75$ K.
This peak can be further analyzed using magnetic field dependent 
resistivity measurements.
\begin{figure}[tb]
\includegraphics[width=18pc]{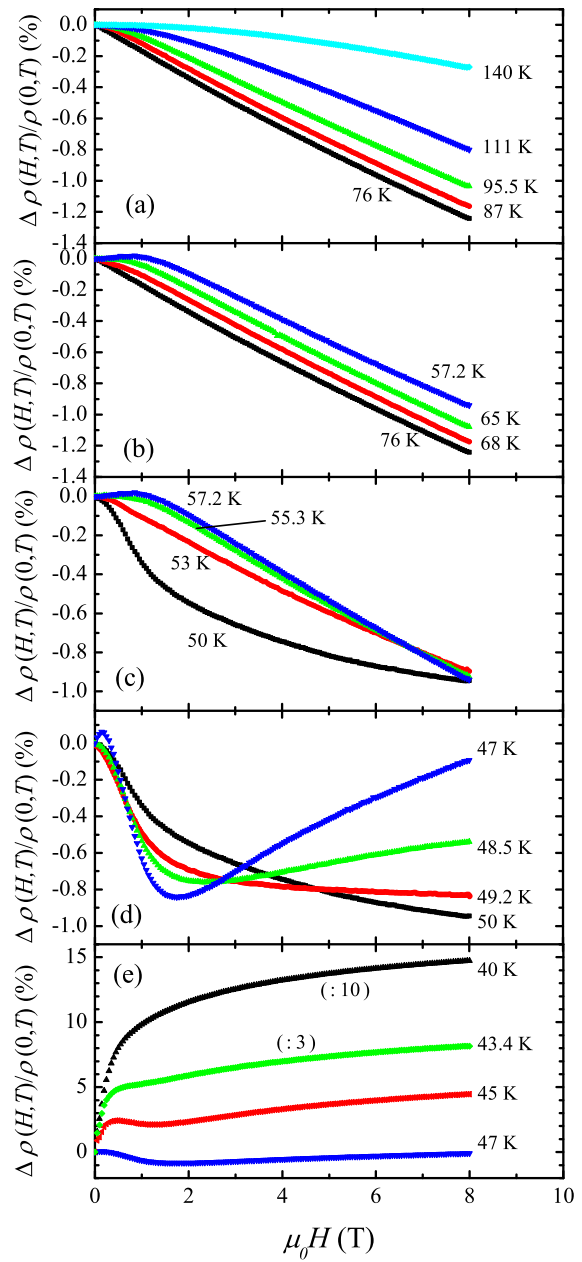}
\caption{(color online) Magnetoresistance of Ru$_{0.8}$Sn$_{0.2}$Sr$_2$GdCu$_2$O$_8$ in various 
temperature ranges. 
The values at 43.4 K and 40 K in (e) are divided by factors 3 and 10, 
respectively.} 
\label{MRall}
\end{figure}
The relative transversal magnetoresistivity 
$[\rho(H,T)-\rho(0,T)]/\rho(0,T)$ is shown in Fig.~\ref{MRall}
in a large temperature range from 140 K down to the superconducting transition. 
The curves are grouped in five subsets according to temperature intervals
in Fig.~\ref{rho}(b) labeled by (a), (b), (c), (d), and (e), exhibiting  
several qualitatively different physical situations.

At temperatures well above $T_m$ [Fig.~\ref{MRall}(a)], the magnetoresistivity is characterized 
by the negative quadratic-in-field behaviour similar to the 
magnetic-field dependence observed in dilute alloys containing uncorrelated 
magnetic impurities \cite{Yosida:57}. 
With decresing temperature the correlations between magnetic Ru ions start 
to play important role and the quadratic-in-field magnetoresistance 
transform into the linear-in-field behaviour.
The same trend can  also be seen for temperatures below $T_m$ 
[Fig.~\ref{MRall}(b)], but not too close to $T_c$.
Not surprisingly, the same linear-in-field behaviour was already observed 
in undoped and in La-doped samples in the vicinity of $T_m$. 
Just below $T_m$, superposed to the negative 
linear magnetoresistivity, a small positive component develops at low fields,
which is related to the AFM ordering of the Ru lattice. 
The same positive contribution is also observed in the longitudinal 
configuration (i.e. $\mathbf H \parallel \mathbf I$).
Finally, it should be noticed that the negative magnetoresistivity of
Figs.~\ref{MRall}(a) and \ref{MRall}(b) does not show any sign of 
saturation up to $\mu_0 H=8\rm \, T$.

Figures~\ref{MRall}(c) and \ref{MRall}(d) reveal a relatively complicated 
behaviour of the magnetoresistivity at temperatures between 57 and 47 K.
The positive AFM contribution decreases rapidly
and vanishes below 53 K. Below 49 K one observes positive contribution, 
presumably related to the SC fluctuations.
The resulting magnetoresistivity, which is the sum of the positive 
SC and  negative FM contribution, 
yields a local minimum.
Finally, below $T=47$ K, the positive  contribution 
to the magnetoresistivity related to the SC ordering of the
conducting layers starts to dominate.
The sharp increase of the magnitude of the relative magnetoresistivity 
in this temperature region, shown  in Fig.~\ref{MRall}(e), 
reflects the sharp decrease of $\rho(0,T)$ in Fig.~\ref{rho}(a).

\begin{figure}[tb]
\includegraphics[width=16pc]{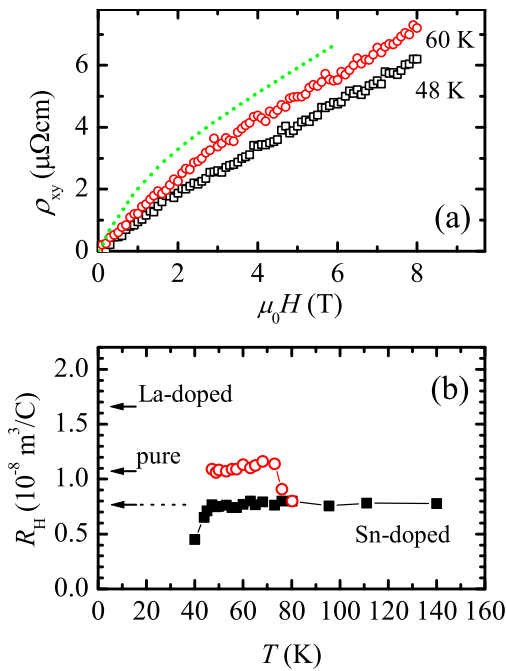}
\caption{(color online) (a) Magnetic field dependence of the Hall resistivity in 
Ru$_{0.8}$Sn$_{0.2}$Sr$_2$GdCu$_2$O$_8$ at $T=60$ K (circles) 
and $T=48$ K (squares). 
The Hall resistivity of the undoped RuSr$_2$GdCu$_2$O$_8$ at $T=124$ K 
is shown by the dotted line for comparison \cite{Pozek:02}. 
(b) Temperature dependence of the low- (open circles) and high-field 
(full squares) Hall coefficients of Ru$_{0.8}$Sn$_{0.2}$Sr$_2$GdCu$_2$O$_8$. 
The averaged high-field Hall coefficients of the pure RuSr$_2$GdCu$_2$O$_8$ 
and the La-doped RuSr$_{1.9}$La$_{0.1}$GdCu$_2$O$_8$
are indicated by the arrows for comparison.} 
\label{Hall}
\end{figure}

So far we have established two seemingly opposed observations:
the resisitivity of the Sn-doped sample is lowered with respect to the pure compound, 
while the electron relaxation rates (proportional to the resisitivity)  
increases with the Sn concentration $x$ (as estimated from the $x$ dependence of $B_{c2}$).
In order to resolve the problem, we have  measured the  Hall resistance 
in which the relaxation rates cancell out in the first approximation, 
allowing in this way the estimation of 
the $x$ dependence of the effective number of charge carriers.

The Hall resistivity of Ru$_{0.8}$Sn$_{0.2}$Sr$_2$GdCu$_2$O$_8$
as a function of magnetic field is shown in 
Fig.~\ref{Hall}(a), for two temperatures below $T_m$, and
compared to the results of the undoped sample \cite{Pozek:02}. 
A nonlinear increase of the Hall resistivity, which is 
the combination of the ordinary and extraordinary Hall contributions, and is
a characteristic of magnetic metals, is observed at low fields. 
At high applied fields, a linear increase dominates,
representing  the ordinary Hall contribution only.
The magnitude of the extraordinary contribution is smaller 
than that in the undoped sample [dotted line in Fig.~\ref{Hall}(a)]. 

The average values of the Hall coefficient $R_H=d\rho _{xy}/d(\mu_0H)$  
are further plotted in Fig.~\ref{Hall}(b) for low and high fields. 
Both, the low- and high-field slopes are smaller than for the pure compound. 
In particular, the absolute value of the ordinary $R_H$ 
is roughly 30\% smaller than in the pure compound, indicating 
a substantial increase of the effective number of charge carriers 
with Sn doping.
Before discussing these  data in more detail (Sec. IV B),
it is useful to recall the prediction for $R_H$ of the single-component 
free-hole model with the carrier concentration  $n_{\rm Cu} = 2p/V_0$,
which is found to explain well the low-temperature Hall measurements in
La$_{2-x}$Sr$_x$CuO$_4$ in the wide doping range $0.02 \le p \le 0.25$
\cite{Ando:04}.
With $V_0 \approx 170.5$ \AA $^3$ being  the primitive cell volume of Ru1212Gd, 
the result is $R_H = 1.065\cdot  10^{-9}/(2p) $ m$^3/$C. The measured high-field
Hall coefficients indicated in Fig.~\ref{Hall}(b) lead to $p = 0.05$, $0.07$ and $0.03$
for the pure, Sn-doped and La-doped samples, respectively.

\section{Discussion}
\label{discussion}
%

In the following, we shall first show that the main changes 
in the zero-field resistivity, and in the (high-field) ordinary 
Hall coefficent with Sn doping (which amounts to several tens of percent
according to the inset of Fig.~\ref{rho}(a) and Fig.~\ref{Hall}),
can be explained by using a single-component model for the conductivity 
tensor (subsection \ref{ordinary}).
Within this model, only CuO$_2$ layers contribute to the observed conductivity, while
 the direct contributions of the RuO$_2$ layers to the total conductivity
are small and can be safely neglected. 
Then, we shall argue, in subsection \ref{magnetic}, that
this single-component model fails to account for the observed 
contribution to the 
magnetoresistivity  (changes of the order of a few percent 
in Fig.~\ref{rho}), and the (low-field) Hall coefficient.
In order to successfully explain these experimental features, 
one has to go beyond the single-component model
and to take into account a presumably small number of itinerant electrons 
in the RuO$_2$ layers.

The motivation for such two-step analysis of our data 
comes from the results of the zero-field NMR (NQR) experiments in the pure 
Ru1212Gd \cite{Tokunaga:01,Kumagai:01}
and the related band structure calculations \cite{Nakamura:01}.
The number of electrons in RuO$_2$ layers is equal to the number 
of holes doped into the CuO$_2$ plane.
Naively, Ru$^{4+}$  ion can be visualized as a composition  
of Ru$^{5+}$ ion and one extra electron.
In this picture, the basic question is as to which extent this extra electron is localized 
to the parent  Ru$^{5+}$ ion.
In Ru-NQR experiments, two well-distinguished  signals are found, corresponding to
Ru$^{5+}$ with the spin state $S= 3/2$ and Ru$^{4+}$ 
with the spin state $S= 1$.
These results suggest that most of the electrons are localized
(leading to Ru$^{4+}$ with the spin state $S=1$),
which means that the number of itinerant (delocalized) electrons
in the RuO$_2$ layers is small.
In NQR, it is also found that the magnetic order in the  RuO$_2$ layers coexists with 
the SC order in the CuO$_2$ planes down to $T=1.4$ K.
Finally, the NQR frequency $\nu^{63}_{Q}\approx 30 \rm \, MHz$ 
measured on $^{63}$Cu nuclei \cite{Kumagai:01,Kramer:02}, 
which is close to $\nu^{63}_{Q}$ measured 
in the underdoped YBa$_2$Cu$_3$O$_{6+x}$ compounds \cite{Yasuoka:89}, 
puts Ru1212$R$ systems into the high-$T_c$ superconductors
with an intermediate copper-oxygen hybridization and, consequently, 
an intermediate value of $T_c$ \cite{KupcicEFG}. 
The band structure calculations  \cite{Nakamura:01}, on the other hand, 
suggest the itinerant character of electrons in the RuO$_2$ layers, 
rather than the localized character, but 
all other conclusions of Ref. \cite{Nakamura:01} 
are consistent with the NQR observations.
Namely, there are  three bands crossing $E_F$, 
the first two are related to two CuO$_2$ layers, while  the third one 
is related to the RuO$_2$ layer.
The magnetic ordering and the calculated magnitude of the magnetic moments
of Ru ions are also consistent with the  local magnetic fields 
measured by NQR \cite{Tokunaga:01,Kumagai:01}.

In the present analysis, the splitting between two hole-like CuO$_2$ 
bands is neglected and the AFM fluctuations in the CuO$_2$ planes
are taken into account \cite{KupcicRaman}. 
This gives the effective concentration of charge carriers (holes) 
in the CuO$_2$ layers $n_{\rm Cu}$ to be proportional to $p$
($n_{\rm Cu} \approx 2p/V_0$) 
rather than to $1+p$, and with the band mass close to the free electron mass.
Thus, $n_{\rm Cu}$ nearly measures the hole concentration in the CuO$_2$ bands
with respect to the half-filling.
The effective concentration of itinerant electrons in the RuO$_2$ layers
$n_{Ru}$ is assumed to be small in comparison with the value
$2p/V_0$ characterizing the completely delocalized case.

\subsection{Resistivity and ordinary Hall coefficient}
\label{ordinary}

In the presence of the magnetic background  of Ru$^{5+}$/Ru$^{4+}$ ions 
and external magnetic fields the conductivity tensor in question is given by
\cite{Ziman,KupcicRaman}
\begin{eqnarray}
\sigma_{xx} &=& \sigma_{yy} = \frac{e^2}{m}\big[\tau_{\rm Cu}
\tilde{n}^{\rm Cu}_{xx} 
+ \tau_{\rm Ru}\tilde{n}^{\rm Ru}_{xx} \big],
\nonumber \\
\sigma_{xy} &=& -\sigma_{yx} = \frac{e^2}{m}\omega _0 
\big[\tau _{\rm Cu}^2 \tilde{n}^{\rm Cu}_{xy}  
 + \tau _{\rm Ru}^2 \tilde{n}^{\rm Ru}_{xy} \big]
 + \sigma_{xy}^{\rm ex}.
\label{eq1}
\end{eqnarray}
The first term in the off-diagonal component $\sigma_{xy}$ 
comes from the Lorentz force and is proportional to $H$
($\omega _0$ is the cyclotron frequency).
The second term, 
$\sigma_{xy}^{\rm ex}$, includes all extraordinary contributions
and is proportional to the magnetization of the Ru lattice,
which means that the dependence on $H$ is given in terms 
of the Brillouin function $B_S(g \mu_0\beta SH)$.
According to Ref.~\cite{Nakamura:01}, the contributions of two bands in 
Eqs.~(\ref{eq1}) can be regarded as decoupled (i.e. the hybridization between 
the CuO$_2$ and RuO$_2$ bands is negligible).
The structure of the effective numbers $\tilde{n}^{i}_{\alpha \beta}$,
$i =$ Cu, Ru and $\alpha, \beta = x, y$, for this case
is given in the Appendix.
In the absence of magnetic fields, for example,  we obtain
$\tilde{n}^{\rm Cu}_{xx}\approx \tilde{n}^{\rm Cu}_{xy} \approx n_{\rm Cu}$ and 
$\tilde{n}^{\rm Ru}_{xx}\approx - \tilde{n}^{\rm Ru}_{xy} \approx n_{\rm Ru}$,  with
$n_{\rm Cu}$ and $ n_{\rm Ru}$ defined above.
$\tau_{\rm Cu}$ and $\tau_{\rm Ru}$ are, respectively,  
the relaxation times of holes and electrons averaged over two spin projections
(see Appendix).
In the absence of magnetic fields, it is reasonable to assume that not only 
$\tilde{n}^{\rm Ru}_{\alpha \beta}$ is small,
when compared to $\tilde{n}^{\rm Cu}_{\alpha \beta}$, but also
the relaxation  rate in the RuO$_2$ layers $1/\tau_{\rm Ru}$ is much larger 
than $1/\tau_{\rm Cu}$,  due to the presence of magnetic ions and the 
substitutional impurities in the RuO$_2$ layers. 
Thus, $\sigma_{\alpha \beta}^{\rm Cu} \gg \sigma_{\alpha \beta}^{\rm Ru}$
for $H=0$.
But, as mentioned above, the main effects in the magnetoresistivity
are expected to come from the field dependence of 
$\sigma_{xx}^{\rm Ru}$ [labeled hereafter by 
$\delta \sigma_{xx}^{\rm Ru} = (e^2/m) \delta \tau_{\rm Ru} 
\tilde{n}^{\rm Ru}_{xx}$].

For weak magnetic fields, the general expressions for the Hall resistivity 
and magnetoresistivity
\begin{eqnarray}
\rho_{xy} &=& \frac{\sigma_{xy}}{\sigma_{xx} \sigma_{yy} + \sigma_{xy}^2},
\nonumber \\
\rho_{xx} &=& \frac{\sigma_{xx}}{\sigma_{xx} \sigma_{yy} + \sigma_{xy}^2}
\label{eq2}
\end{eqnarray}
lead to 
\begin{eqnarray}
\rho_{xy}  &\approx& \frac{\sigma_{xy}^{\rm Cu} + \sigma_{xy}^{\rm Ru}}{
\big(\sigma_{xx}^{\rm Cu} + \sigma_{xx}^{\rm Ru} \big)^2},
\nonumber \\
\rho_{xx} &\approx& \frac{1}{\sigma_{xx}^{\rm Cu} + \sigma_{xx}^{\rm Ru}}
\label{eq3}
\end{eqnarray}
i.e.
\begin{eqnarray}
\rho_{xy}  &\approx& \frac{m }{e^2}
\frac{\omega_0[\tau_{\rm Cu}^2 n_{\rm Cu} - \tau _{\rm Ru}^2 n_{\rm Ru}]
+(m/e^2)\sigma_{xy}^{\rm ex; Ru}}{
\big(\tau_{\rm Cu} n_{\rm Cu} + \tau_{\rm Ru} n_{\rm Ru} \big)^2},
\nonumber \\
\rho_{xx}  &\approx& \frac{m}{e^2}
\frac{1}{\tau_{\rm Cu} n_{\rm Cu} + \tau_{\rm Ru} n_{\rm Ru}}.
\label{eq4}
\end{eqnarray}
%
The Hall coefficient is given by $R_H=d\rho _{xy}/d(\mu_0H)$.
The anomalous contributions (the negative magnetoresistivity and 
the extraordinary Hall coefficent) are hidden here in
$\delta \sigma_{xx}^{\rm Ru}$  and  $\sigma_{xy}^{\rm ex; Ru}$, and we put 
$\delta \sigma_{xx}^{\rm Ru} = \sigma_{xy}^{\rm ex; Ru} = 0$ 
in the rest of this subsection.

In the Sn doped samples, the replacement of any Ru atom by Sn introduces 
one extra hole in the system, which  is  redistributed between RuO$_2$ 
and CuO$_2$ layers.
The part of the extra charge which remains in the 
RuO$_2$ layers decreases the effective number  of itinerant electrons 
$n_{\rm Ru}$ ($\Delta n_{\rm Ru} <0$), thus increasing both the resistivity and 
Hall resistivity. One may say that it gives rise to the conversion of Ru$^{4+}$ into 
Ru$^{5+}$.  The part of the extra charge transferred into the CuO$_2$ layers 
increases the effective number $n_{\rm Cu}$ ($\Delta n_{\rm Cu} >0)$.
The experimentally determined decrease of the normal-state 
resistivity  and the (high-field) Hall coefficent with the Sn doping
(Figs.~\ref{rho} and \ref{Hall}) shows that a substantial portion of
the doped holes are indeed transferred to the CuO$_2$ layers and 
that the small changes in already small $n_{\rm Ru}$ can be neglected.
This scenario is expected to hold for not too large number of Sn impurities
($x \le 0.3$).
Our results should be contrasted to the Sn doping study of 
McCrone et. al (Fig. 2 in Ref. \cite{McCrone:03}), where the increase of 
resistivity and the Hall coefficient was observed for $x=0.2$.

As mentioned above, the correlation between $n_{\rm Cu}$ 
and the superconducting critical temperature $T_c$ is expected to be similar 
to the correlations found in the isostructural YBa$_2$Cu$_3$O$_{7-x}$ compounds.
The main difference is that a very complicated role of the CuO chains 
in supporting superconductivity in the CuO$_2$ layers is played  here 
by the magnetically ordered RuO$_2$ layers. 
Indeed, the dependence of $T_c$ on the Sn doping $x$ (shown in Fig.~3) is
similar to the observation in the underdoped YBa$_2$Cu$_3$O$_{7-x}$ compounds
($T_c$ is nearly constant in a wide range of $x$, $T_c\approx 60$ K in
YBa$_2$Cu$_3$O$_{7-x}$).

Similar scenario holds also for other Ru1212$R$ systems.
For example, the replacement of Sr$^{2+}$ by La$^{3+}$ increases 
the total number of conduction electrons in the system. 
A substantial part of the doped electrons are transferred to the CuO$_2$ 
layers, reducing the effective number $n_{\rm Cu}$ below the critical value 
required for superconductivity, in agreement with experiments.
In RuSr$_{1.9}$La$_{0.1}$GdCu$_2$O$_8$, the high temperature resistivity and
the high-field Hall resistivity are found to be increased by roughly 60-70\%
 with respect to the pure compound, 
and the superconductivity is completely suppressed \cite{Pozek:07}. 
%
%
The same  effect of reducing the effective number $n_{\rm Cu}$ 
is obtained by the partial replacement of Ru ions 
by Nb \cite{McCrone:03,McLaughlin:01}.

\subsection{Magnetic properties}
\label{magnetic}

The magnetoresistivity data of Fig.~\ref{MRall} give us the indirect way 
to study the magnetic properties of the Ru lattice.
In the model described above, we can write
\begin{eqnarray}
\frac{\Delta \rho_{xx}}{\rho_{xx}(0)}=
\frac{\rho_{xx} (H) - \rho_{xx}(0)}{\rho_{xx}(0)}&\approx& 
\frac{\sigma_{xx}^{\rm Ru}(0)}{\sigma_{xx}(0)}
\frac{\delta \rho_{xx}^{\rm Ru}}{ \rho_{xx}^{\rm Ru}(0)},
\label{eq5}
\end{eqnarray}
with $\sigma_{xx}(0)$ and $\sigma_{xx}^{\rm Ru}(0)$ being, respectively,
the zero-field total conductivity and the zero-field conductivity of the
RuO$_2$ subsystem.
The first factor is nearly proportional to 
$n_{\rm Ru}/(n_{\rm Cu} + n_{\rm Ru})$, and represents an enhancement factor
dependent on Sn doping.
It grows with decreasing $n_{\rm Cu}$, as can be easily seen from
our data taken in three compounds: Ru$_{0.8}$Sn$_{0.2}$Sr$_2$GdCu$_2$O$_8$,
Ru$_1$Sr$_2$GdCu$_2$O$_8$, and  Ru$_{0.9}$La$_{0.1}$Sr$_2$GdCu$_2$O$_8$.
The second factor in Eq.~(\ref{eq5}), $\delta \rho_{xx}^{\rm Ru}/ \rho_{xx}^{\rm Ru}(0)
\approx -\delta \sigma_{xx}^{\rm Ru}/\sigma_{xx}^{\rm Ru}(0)$,
is the magnetoresistivity of the 
isolated RuO$_2$ subsystem, and it is expected to be similar 
to that of magnetic metals \cite{Yosida:57,Fert:77,Kataoka:01}.
Precisely, the negative quadratic-in-field magnetoresistivity observed at
temperatures well above $T_m$ has to be contrasted to the magnetoresistivity of 
common high-$T_c$ superconductors which is very small and positive
\cite{Harris:95}.
This high temperature behaviour can be understood as a clear evidence
of the exchange interaction between itinerant electrons in the RuO$_2$
layers and the localized Ru spins \cite{Yosida:57}.
The AFM interactions between localized spins become important at low enough
temperatures leading to the linear-in-field behaviour with maximal slope at
$T \approx T_m$ [Figs.~\ref{MRall}(a)-(c)].
We recall that exactly the same temperature evolution of the magnetoresistivity 
was found in Ref.~\cite{Kataoka:01} for small concentration 
of conduction electrons and low stability of the FM state 
of the localized spins.
Hence, we believe that expression (\ref{eq5}) 
includes all relevant processes in the magnetoresistivity of ruthenate 
cupartes in the nonsuperconducting temperature range.

Taking into account that $\tau_{\rm Ru}n_{\rm Ru} \ll \tau_{\rm Cu}n_{\rm Cu}$, 
one can rewrite Eq.~(\ref{eq5}) in the form:
\begin{eqnarray}
\frac{\delta \rho_{xx}^{\rm Ru}}{\rho_{xx}^{\rm Ru}(0)}
\approx
\frac{\tau_{\rm Cu} n_{\rm Cu}}{\tau_{\rm Ru} n_{\rm Ru}}
\frac{\Delta \rho_{xx}}{\rho_{xx}(0)}
\approx 
-\frac{\tau_{\rm Cu}}{\tau_{\rm Ru} V_0 n_{\rm Ru}} \frac{V_0}{R_H e}
\frac{\Delta \rho_{xx}}{\rho_{xx}(0)}
,
\label{eq6}
\end{eqnarray}
\begin{figure}[tb]
\includegraphics[width=16pc]{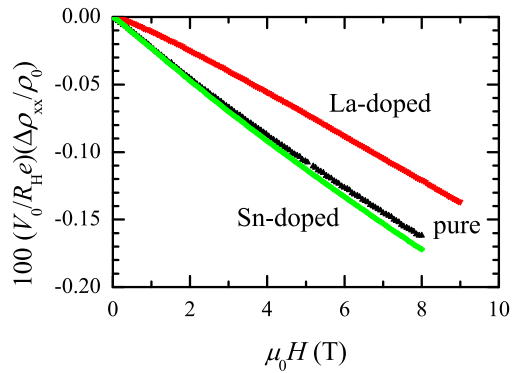}
\caption{(color online) Field dependence of the quantity $V_0 /(R_H e)
(\Delta \rho_{xx}/\rho_{xx}(0))$ for the pure (RuSr$_2$GdCu$_2$O$_8$), 
Sn-doped (Ru$_{0.8}$Sn$_{0.2}$Sr$_2$GdCu$_2$O$_8$) and 
La-doped (RuSr$_{1.9}$La$_{0.1}$GdCu$_2$O$_8$) samples 
at their respective magnetic ordering temperatures.} 
\label{colossal}
\end{figure}
where $\Delta \rho_{xx}=\rho_{xx} (H) - \rho_{xx}(0)$.
Figure \ref{colossal} shows the field dependence of the quantity $V_0 /(R_H e) 
(\Delta \rho_{xx}/\rho_{xx}(0))$ for the three aforementioned samples 
at their respective magnetic ordering temperatures.
If one assumes, based on the previous subsection, that $V_0 n_{Ru}<0.01$ and 
$\tau_{Ru}<\tau_{Cu}$, the magnetoconductivity of ruthenium layers is found to 
exceed 18\% at $\mu_0 H=8\rm \, T$.
The magnitude of this effect is comparable to the colossal magnetoresistivity of the 
manganite perovskites.

\section{Conclusions}

In conclusion, a microwave study of Sn doping in Ru1212 samples has shown that 
magnetic ordering is strongly influenced by introduction of tin into RuO$_2$ layers, 
while the superconducting critical temperature remains practically unchanged. 

The magnetic field dependence of the microwave absorption has shown that the upper 
critical field in doped samples is increased, probably due to the increased disorder. 
However, in spite of increased disorder, the dc resistivity measurement in 
Ru$_{0.8}$Sn$_{0.2}$Sr$_2$GdCu$_2$O$_8$ is lower than in the respective pure compound. 
We conclude that it is due to the increased number of holes in the CuO$_2$ layers. 
This conclusion is also confirmed by Hall resistivity measurements presented in this paper. 

In Ru$_{0.8}$Sn$_{0.2}$Sr$_2$GdCu$_2$O$_8$ magnetoresistivity 
develops from quadratic-in-field behavior (characteristic for uncorrelated magnetic impurities)
at high temperatures to linear-in-field behavior close to $T_m\approx 76 \rm\, {K}$, 
indicating significant correlations between magnetic Ru ions.
Below $T_m$, a small positive component of magnetoresistivity, related to the AFM 
ordering of the RuO$_2$ lattice, is superposed to the negative linear magnetoresistivity.
As the temperature is lowered below $53\rm\, {K}$, 
the field-induced ferromagnetic order seems to prevail. 
At the onset of superconductivity ($\approx 48\rm\, {K}$) both, positive 
superconducting and negative ferromagnetic contributions to the same 
magnetoresistivity curve occur.

We have shown that the single component model for the conductivity tensor suffices
to explain the variation of the normal-state resistivity and the high-field Hall coefficient
with doping. The main conductivity channel is through CuO$_2$ layers and the various dopings 
simply change the number of conducting holes in these layers.
However, the low-field Hall resistivity and the magnetoresistivity provide evidence that there is 
a small conduction in magnetically ordered RuO$_2$ layers. 
Quantitative analysis of the observed
negative magnetoresistance, leads to the conclusion that small number 
of delocalized electrons display a colossal magnetoresistance
which
exceeds 18\%. This is a remarkable result.

\section*{Acknowledgments}

We acknowledge funding support from the Croatian Ministry of Science, 
Education and Sports (projects no. 119-1191458-1022 ``Microwave Investigations 
of New Materials'', no. 119-1191458-0512 
``Low-dimensional strongly correlated conducting systems'' 
and no. 119-1191458-1023 ``Systems with spatial and dimensional constraints: 
correlations and spin effects''), the 
New Zealand Marsden Fund, the New Zealand Foundation for Research Science 
and Technology, and the Alexander von Humboldt Foundation.

\appendix

\section{Conductivity tensor}

In the multiband models with the interband hybridization negligible 
(Ru1212$R$ systems are the example),
the elements of the conductivity tensor are given by the sum of the 
intraband terms $\sigma_{\alpha \beta} = \sum_i \sigma^i_{\alpha \beta}$
($i =$ Cu, Ru in the present case).
The  presence of  magnetic background and  external magnetic fields
makes, in the first place, that the relaxation rates of conduction electrons
$\Gamma_{i\sigma} ({\bf k}) = 1/ \tau_{i\sigma}({\bf k})$ depend on spin,
and also introduces the extraordinary contributions to the 
off-diagonal elements of the conductivity tensor $\sigma_{xy}^{{\rm ex}:i}$,
as discussed in the main text.
In the first approximation (the case of noninteracting magnetic impurities
\cite{Yosida:57}) we can write $\Gamma_{i\sigma} ({\bf k}) \approx
\Gamma_{i} \pm \alpha_i M_{z}$, where $\alpha_i$ is a constant and 
$M_{z}$ is the total magnetization of the RuO$_2$ layer).
Thus, in the Ru1212$R$ systems, at temperatures well above $T_m$, we have 
$\sigma^i_{xx} = \sigma^i_{yy} =(e^2/m) \tau_i \tilde{n}_{xx}^i$ and
$\sigma^i_{yx} = -\sigma^i_{xy} =(e^2/m) \omega_0 \tau_i^2 \tilde{n}_{xy}^i
+ \sigma_{yx}^{{\rm ex}:i}$, 
with $\tau_i \tilde{n}_{xx}^i$ and $\tau_i^2 \tilde{n}_{xy}^i$
given by \cite{Ziman,KupcicRaman}
\begin{eqnarray}
\tau_i \tilde{n}_{xx}^i &\approx&   \frac{1}{V} \sum_{{\bf k} \sigma}
\frac{ m [v_x^i ({\bf k})]^2 \tau_{i \sigma} ({\bf k})}{
1 + \tau_{i \sigma}^2 ({\bf k}) \omega_{i0}^2 ({\bf k})}
(-)\frac{\partial f_{i \sigma} ({\bf k})}{\partial \varepsilon_{i \sigma} ({\bf k})},
\nonumber \\ 
\tau_i^2 \tilde{n}_{xy}^i &\approx& 
\frac{1}{V} \sum_{{\bf k} \sigma}
\frac{ m v_x^i ({\bf k})\big[v_x^i ({\bf k}) \gamma_{yy}^i ({\bf k})
- v_y^i ({\bf k})\gamma_{xy}^i ({\bf k}) \big]
}{1 + \tau_{i \sigma}^2 ({\bf k}) \omega_{i0}^2 ({\bf k})}
\nonumber \\ 
&& \times 
\tau_{i \sigma}^2 ({\bf k})(-)\frac{\partial 
f_{i \sigma} ({\bf k})}{\partial \varepsilon_{i \sigma} ({\bf k})}.
\label{eqA1}
\end{eqnarray}
$v_x^i ({\bf k})$ is the electron group velocity, 
$\omega_{i0}^2 ({\bf k}) = 
\gamma_{xx}^i ({\bf k}) \gamma_{yy}^i ({\bf k}) \omega_0^2$,
and $\gamma_{\alpha \beta}^i ({\bf k})$ are the elements of 
the inverse effective-mass tensor [$\gamma_{xx}^i ({\bf k}=0) =
m/m^*_i$].

If  magnetic fields are weak  and the exchange interaction between 
the conduction holes and localized Ru spins is negligible, 
the field dependence in the denominators in (\ref{eqA1})
can be neglected, and the sum over spin projections gives 
for the averaged relaxation times of itinerant electrons the expression
$\sum_{\sigma} \tau_{{\rm Ru} \sigma}  ({\bf k}) \approx
2 \Gamma^0_{\rm Ru}/\big[(\Gamma^0_{\rm Ru})^2 - \alpha^2_{\rm Ru}
(M_z)^2 \big]$, resulting in  
$\delta \sigma_{xx}^{\rm Ru} \propto H^2$ \cite{Yosida:57}.
According to Eq.~(\ref{eq5}), this also leads to the total magnetoresistivity
which is negative and quadratic-in-field.
The effects of interactions between localized spins 
on the conductivity tensor at different temperatures 
are expected to be similar to that found in Ref.~\cite{Kataoka:01}.


\begin{thebibliography}{00}

\bibitem{Nachtrab:06} 
T. Nachtrab, C. Bernhard, L. Chengtian, D. Koelle, and R. Kleiner, 
Comptes Rendus Physique {\bf 7}, 68 (2006).

\bibitem{McCrone:03} 
J.~E.~McCrone, J.~L.~Tallon, J.~R.~Cooper, A.~C.~McLaughlin, J.~P.~Attfield, 
and C.~Bernhard, 
Phys. Rev. B {\bf 68}, 064514  (2003).

\bibitem{Klamut:01}
P. W. Klamut, B. Dabrowski, S. Kolesnik, M. Maxwell, and J. Mais, 
Phys. Rev. B {\bf 63}, 224512 (2001). 

\bibitem{McLaughlin:01} 
A. C. Mclaughlin, V. Janowitz, J. A. McAllister, and J. P. Attfield, 
J. Mater. Chem {\bf 11}, 173 (2001).

\bibitem{Williams:03} 
G.~V.~M.~Williams, H.~K.~Lee, and S.~Kr\"{a}mer,
Phys. Rev. B {\bf 67}, 104514 (2003).

\bibitem{Hassen:06} 
A. Hassen and P. Mandal, 
Supercond. Sci. Technol. {\bf 19}, 902 (2006).

\bibitem{McLaughlin:99} 
A. C. Mclaughlin and J. P. Attfield, 
Phys. Rev. B {\bf 60}, 14605 (1999). 

\bibitem{Malo:00} 
S. Malo, K. Donggeun, J. T. Rijssenbeek, A. Maignan, D. Pelloquin, V. P. Dravid, 
and K. R. Poeppelmeier, Int. J. Inorg. Mater.{\bf 2}, 601 (2000).

\bibitem{Hassen:03} 
A. Hassen, J. Hemberger, A. Loidl, and A. Krimmel, 
Physica C {\bf 400}, 71 (2003).

\bibitem{Yamada:04} 
Y. Yamada, H. Hamada, M. Shimada, S. Kubo, and A. Matsushita, 
J. Magn. Magn. Mater. {\bf 272-276}, e173 (2004).

\bibitem{Steiger:07} 
M. Steiger, C. Kongmark, F. Rueckert, L. Harding, and M. S. Torikachvili, 
Physica C  {\bf 453}, 24 (2007).

\bibitem{Klamut:01a} 
P. W. Klamut, B. Dabrowski, J. Mais, and M. Maxwell,  
Physica C {\bf 350}, 24 (2001).

\bibitem{Papageorgiou:07}
T. P. Papageorgiou, E. Casini, Y. Skourski, T. Herrmannsd\"orfer, J. Freudenberger, H. F. Braun, and J. Wosnitza,
Phys. Rev. B {\bf 75}, 104513 (2007).

\bibitem{Tokunaga:01}
Y.~Tokunaga, H.~Kotegawa, K.~Ishida, Y.~Kitaoka, H.~Takagiwa, and J.~Akimitsu, 
Phys. Rev. Lett. {\bf 86}, 5767 (2001).

\bibitem{Kumagai:01}
K. Kumagai, S. Takada and Y. Furukawa,
Phys. Rev. B {\bf 63}, 180509 (2001).

\bibitem{Pozek:02}
M.~Po\v{z}ek, A.~Dul\v{c}i\'{c}, D.~Paar, A.~Hamzi\'{c}, M.~Basleti\'{c}, E.~Tafra, 
G.~V.~M.~Williams, and S.~Kr\"{a}mer, 
Phys. Rev. B {\bf 65}, 174514  (2002).

\bibitem{Pozek:07}
M.~Po\v{z}ek, A.~Dul\v{c}i\'{c}, A.~Hamzi\'{c}, M.~Basleti\'{c}, E.~Tafra, 
G.~V.~M.~Williams, and S.~Kr\"{a}mer, 
Eur. Phys. J. B {\bf 57}, 1 (2007).

\bibitem{Nebendahl:01}
B.~Nebendahl, D.-N.~Peligrad, M.~Po\v{z}ek, A.~Dul\v{c}i\'{c}, and M.~Mehring, 
Rev. Sci. Instrum. {\bf 72}, 1876 (2001).

\bibitem{Janjusevic:06}
D. Janju\v{s}evi\'{c}, M. Grbi\'c, M. Po\v zek, A.~Dul\v{c}i\'{c}, D.~Paar, B. Nebendahl, and T. Wagner,
Phys. Rev. B {\bf 74}, 104501 (2006).

\bibitem{Yosida:57}
K. Yosida,
Phys. Rev.  {\bf 107}, 396 (1957).

\bibitem{Ando:04}
Y. Ando, Y. Kurita, S. Komiya, S. Ono, and K. Segawa,
Phys. Rev. Lett. {\bf 92}, 197001 (2004).

\bibitem{Nakamura:01}
K. Nakamura, K. T. Park, A. J. Freeman, and J. D. Jorgensen, 
Phys. Rev. B {\bf 63}, 024507  (2001).

\bibitem{Kramer:02}
S. Kr\"{a}mer and G. V. M. Williams,
Physica C {\bf 377}, 282 (2002).

\bibitem{Yasuoka:89}
H. Yasuoka, T. Imai, and T. Shimizu,
{\it Strong Correlations and Superconductivity}, 
Vol. 89 of Springer Serier in Solid State Sciences
(Springer, Berlin, 1989), p. 254.

\bibitem{KupcicEFG}
I. Kup\v{c}i\'{c}, S. Bari\v{s}i\'{c}, and E. Tuti\v{s},
Phys. Rev. B {\bf 57}, 8590 (1998).

\bibitem{KupcicRaman}
I. Kup\v{c}i\'{c} and  S. Bari\v{s}i\'{c},
Phys. Rev. B {\bf 75}, 094508 (2007).

\bibitem{Ziman}
J. M. Ziman, 
{\it Electrons and Phonons} 
(Oxford University Press, London, 1972).

\bibitem{Fert:77}
A. Fert, R. Asomoza, D. H. Sanchez, D. Spanjaard, and A. Friederich,
Phys. Rev. B {\bf 16}, 5040 (1977).

\bibitem{Kataoka:01}
M. Kataoka, 
Phys. Rev. B {\bf 63}, 134435  (2001).

\bibitem{Harris:95}
J. M. Harris, Y. F. Yan, P. Matl, N. P. Ong, P. W. Anderson, T. Kimura, and K. Kitazawa,
Phys. Rev. Lett. {\bf 75}, 1391 (1995).


\end{thebibliography}
\end{document}